# Securing Software as a Service Model of Cloud Computing: Issues and Solutions

Rashmi [1], Dr.G.Sahoo[2], Dr.S.Mehfuz[3]

[1,2]Birla Institute of Technology, Mesra, Ranchi, Jharkhand, India[3]Jamia Milia Islamia,Delhi,India

*ABSTRACT*

*Cloud computing, undoubtedly, has become the buzzword in the IT industry today. Looking at the potential impact it has on numerous business applications as well as in our everyday life, it can certainly be said that this disruptive technology is here to stay. Many of the features that make cloud computing attractive, have not just challenged the existing security system, but have also revealed new security issues. This paper provides an insightful analysis of the existing status on cloud computing security issues based on a detailed survey carried by the author. It also makes an attempt to describe the security challenges in Software as a Service (SaaS) model of cloud computing and also endeavors to provide future security research directions.*

## 1.INTRODUCTION

A lot has been written and spoken about Cloud Computing technology, by IT experts, industry and business leaders and independent experts. While some believe it is a disruptive trend representing the next stage in the evolution of the Internet, others believe it is hype, as it uses earlier established computing technologies. So, what exactly is cloud computing? From a user perspective, cloud computing provides a means for acquiring computing services without any need for deep understanding of the underlying technology being used. From an organizational perspective, cloud computing delivers services for consumer and business needs in a simplified way, providing unbounded scale and differentiated quality of service to foster rapid innovation and decision making. According to Gartner [1], cloud computing can be defined as ''a style of computing, where massively scalable IT- enabled capabilities are delivered 'as a service' to external customers using Internet technologies. According to the Seccombe [2] and National Institute of Standards & Technology [3], guidelines for cloud computing, it has four different deployment models namely private, community, public and hybrid as well as three different delivery models that are utilized within a particular deployment model. These delivery models are the SaaS (Software as a Service), PaaS (Platform as a Service) and IaaS (Infrastructure as a Service). These models form the core of the cloud and they exhibit certain key characteristics like on demand self-service, broad network access, resource pooling, measured service and rapid elasticity. Our main area of concern in this paper is the Software as a service (SaaS). model This best-known branch of cloud computing , is a delivery model in which applications are hosted and managed in a service provider's datacenter, paid for on a subscription basis and accessed via a browser over an internet connection. It basically deals with licensing of an application to the customers for use as a service on demand. One good example of SaaS is the Salesforce.com CRM application.

This paper focuses on the issues related to the service delivery model of cloud computing. The paper describes the various security issues of cloud computing with respect to its service delivery model SaaS. The organization of the paper is as follows: Section 2 describes the security issues that are posed by the Software as a Service (SaaS) delivery model. Section 3 lists some of the current solutions which partly target the security challenges posed by the SaaS. Section 4 provides conclusions derived out of the survey undertaken.

## 2. SECURITY ISSUES IN SAAS

In Software as a Service (SaaS) model, the client has to depend on the service provider for proper security measures. The provider must ensure that the multiple users don't get to see each other's data. So, it becomes important to the user to ensure that right security measures are in place and also difficult to get an assurance that the application will be available when needed [4]. While using SaaS model, the cloud customer will, by definition, be substituting new software applications for old ones. Therefore, the focus is not upon portability of applications, but on preserving or enhancing the security functionality provided by the legacy application and achieving a successful data migration [2] The SaaS software vendor may host the application on his own private server or deploy it on a cloud computing infrastructure service provided by a third-party provider (e.g. Amazon, Google, etc.). The use of cloud computing coupled with the 'pay-as-you-go' approach helps the application service provider reduce the investment in infrastructure services and enables it to concentrate on providing better services to the customers. Enterprises today view data and business transactions as strategic and guard them with access control and compliance policies. However, in the SaaS model, enterprise data is stored at the SaaS provider's data center, along with the data of other enterprises. Moreover, if the SaaS provider is leveraging a public cloud computing service, the enterprise data might be stored along with the data of other unrelated SaaS applications. The cloud provider might, additionally, replicate the data at multiple locations across countries for the purposes of maintaining high availability. Most enterprises are familiar with the traditional on- promise model, where the data continues to reside within the enterprise boundary, subject to their policies. Cloud computing providers need to solve the common security challenges being faced by traditional communication systems. At the same time, they also have to deal with other issues inherently introduced by the cloud computing paradigm itself. In the following section, the SaaS security issues have been categorized as traditional and new cloud specific security challenges, for sake of convenience. Figure 1 shows the various security issues in SaaS model which are discussed in the paper.

### 2.1    Traditional Security Challenges

Although the security concerns in traditional communication systems also apply to the cloud, the use of cloud computing introduces new attack vectors that will make attacks either possible or simply easier to carry out. Some of the traditional security issues which also affect the SaaS model have been described below:

#### Authentication and authorization

The authentication and authorization applications for enterprise environments may need to be changed, to work with a safe cloud environment. Forensics tasks may become much more difficult since the investigators may not be able to access system hardware physically. The design proposed by Pratap Murukutla [5] allows user to use a single set of credentials. They have proposed a solution with de-facto standards of open authorization in which there is a trust party auditor which maintains all the credentials and cloud provider can uniquely distinguish one user from other. The model proposed in the literature [6] verifies user authenticity using two-step verification, which is based on password, smartcard and out of band (i.e. strong two factors) authentication. In addition, the scheme also provides mutual authentication, identity management, session key establishment, user privacy and security against many popular attacks; however the formal security proofing hasn't yet been formalized.

## Availability

The availability ensures the reliable and timely access to cloud data or cloud computing resources by the appropriate personnel. The availability of cloud service providers is also a big concern, since if the cloud service is disrupted; it affects more customers than in the traditional model. For instance, the recent disruption of the Amazon cloud service in the year 2011, took down a number of websites including Reddit, Foursquare, and Quora.The SaaS application providers are required to ensure that the systems are running properly when needed and enterprises are provided with services around the clock. This involves making architectural changes at the application and infrastructural levels to add scalability and high availability. Resiliency to hardware/software failures, as well as to denial of service attacks, needs to be built from the ground up within the application. At the same time, an appropriate action plan for business continuity and Disaster Recovery (DR) needs to be considered for any exigencies as per the guidance provided by [2]. This is essential to ensure the safety of the enterprise data while maintaining minimal downtime for the enterprises. With Amazon [7] for instance, the Amazon Web Services (AWS) API end points are hosted on the same Internet-scale, world-class infrastructure that supports the Amazon retail site. Standard Distributed Denial of Service (DDoS) mitigation techniques such as synchronous cookies and connection limiting are used. To further mitigate the effect to potential DDoS attacks, Amazon maintains internal bandwidth that exceeds its provider-supplied Internet bandwidth.

## Data confidentiality

Confidentiality refers to the prevention of intentional or unintentional unauthorized disclosure of information. Confidentiality in cloud system is related to the areas of intellectual property rights, covert channels, traffic analysis, encryption, and inference [10]. Cloud computing involves the sharing or storage of information on remote servers owned or operated by others, while accessing through the Internet or any other connections. Cloud computing services exist in many variations, including data storage sites, video sites, tax preparation sites, personal health record websites and many more. The entire contents of a user's storage device may be stored with a single cloud provider or with multiple cloud providers. Whenever an individual, a business, a government agency, or any other entity shares information in the cloud, privacy or confidentiality questions arise.

## Virtual Machine Security

Although the global adoption of virtualization is a relatively a recent phenomena, threats to the virtualized infrastructure are evolving just as quickly [8]. The hypervisor and virtual machines used in cloud providers may also have vulnerabilities, as exemplified by [14]. Such vulnerabilities represent an even more serious problem in multi-tenant environments, where compromise of even a single virtual machine can affect all users on the same physical server.

Virtualization is one of the main components of a cloud. But this poses major security risks. Ensuring that different instances running on the same physical machine are isolated from each other is a major task of virtualization which is not met completely in today's scenario. The other issue is the control of administrator on host and guest operating systems. Current Virtual Machine Monitor (VMMs do not offer perfect isolation. Many bugs have been found in all popular VMMs that allow escaping from VM. Virtual machine monitor should be 'root secure', meaning that no privilege within the virtualized guest environment permits interference with the host system. Some vulnerability has been found in all virtualization software which can be exploited by malicious, local users to bypass certain security restrictions or gain privileges. For example, the vulnerability of Microsoft Virtual PC and Microsoft Virtual Server could allow a guest operating system user to run code on the host or another guest operating system. Vulnerability in Virtual PC and Virtual Server could allow elevation of privilege. Cloud providers, therefore, might need to reconsider traditional security concerns from different angles.

Figure 1: Security Issues in SaaS

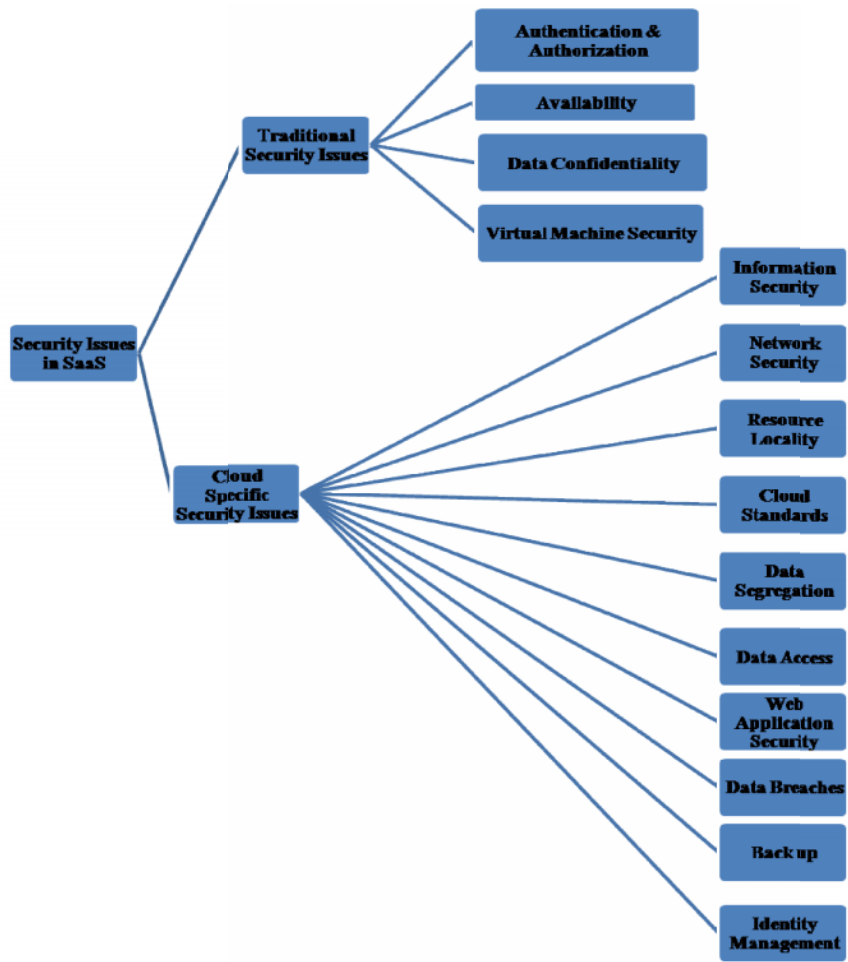

## 2.2    Cloud Specific Security Challenges

**Information  Security**

In a traditional on-premise application deployment model, the sensitive data of each enterprise continues to reside within the enterprise boundary and is subject to its physical, logical and personnel security and access control policies. However, in the SaaS model, the enterprise data is stored outside the enterprise boundary, at the SaaS vendor end. Consequently, the SaaS vendor must adopt additional security checks to ensure data security and prevent breaches due to security vulnerabilities in the application or through malicious employees. This involves the use of strong encryption techniques for data security and fine-grained authorization to control access to data.

In cloud vendors such as Amazon, the Elastic Compute Cloud (EC2) administrators do not have access to customer instances and cannot log into the Guest OS. EC2 Administrators with a business need are required to use their individual cryptographically Strong Secure Shell (SSH) keys to gain access to a host [8]. All such accesses are logged and routinely audited. While the data at rest in Simple Storage Service (S3) is not encrypted by default, users can encrypt their data before it is uploaded to Amazon S3, so that it is not accessed or tampered with by any unauthorized party. Malicious users can exploit weaknesses in the data security model to gain unauthorized access to data.

**Network Security**

In a SaaS deployment model, sensitive data is obtained from the enterprises, processed by the SaaS application and stored at the SaaS vendor end. All data flow over the network needs to be secured in order to prevent leakage of sensitive information. This involves the use of strong network traffic encryption techniques such as Secure Socket Layer (SSL) and the Transport Layer Security (TLS) for security.

In case of Amazon Web Services (AWS), the network layer provides significant protection against traditional network security issues, such as MITM (Man-In-The-Middle) attacks, IP spoofing, port scanning, packet sniffing, etc. For maximum security, Amazon S3 is accessible via SSL encrypted endpoints. The encrypted end points are accessible from both the Internet and from within AmazonEC2, ensuring that data is transferred securely both within AWS and to and from sources outside of AWS [9]. However, malicious users can exploit weaknesses in network security configuration to sniff network packets.

**Resource Localty**

In a SaaS model of a cloud environment, the end-users use the services provided by the cloud providers without knowing exactly where the resources for such services are located, possibly in other legislative domains. This poses a potential problem when disputes happen, which is sometimes beyond the control of cloud providers. Due to compliance and data privacy laws in various countries, locality of data is of utmost importance in much enterprise architecture [15]. The European Union has issued a Directive 95/46/EC to protect the user privacy at all costs [16]. The directive prohibits transfers of personal data to countries which do not ensure an adequate level of protection. For example, the recent Dropbox users have to agree to the ''Terms of Service'' which grant the providers the right to disclose user information in compliance with laws and law enforcement requests [17]. In addition to the issue of local laws, there's also the question of under whose jurisdiction the data falls, when an investigation occurs. A secure SaaS model must be capable of providing reliability to the customer on the location of the data of the consumer.

**Cloud standards**

To achieve interoperability among clouds and to increase their stability and security, cloud standards are needed across different standard developing organizations. For example, the current storage services by a cloud provider may be incompatible with those of other provider. In order to keep their customers, cloud providers may introduce so called ''sticky services'' which create difficulty for the users if they want to migrate from one provider to the other, e.g., Amazon's S3 is incompatible with IBM's Blue Cloud or Google storage. There are currently a large number of standards bodies with different interests, e.g. IEEE Cloud Computing Standard Study Group [18], ITU Cloud Computing Focus Group [19], Cloud Security Alliance (CSA), Distributed Management Task Force [20], Storage Networking Industry Association [21], Open Grid Forum [22], Open Cloud Consortium [23] and Organization for the Advancement of Structured

Information Standards [24], and so forth. To promote the wide use of cloud computing, these standards bodies need to sit down and work together to establish common standards. Possible ''Inter-cloud'' standards in the following domains are needed to increase cloud interoperability and free data movement among clouds:

– Network architecture,
– Data format,
– Metering and billing,
– Quality of Service,
– Resource provisioning,
– Security, identity management and privacy.

As stated, there are many general computing standards that may be reused in the cloud, but for the moment, there are to our knowledge no dedicated cloud standards. This may add to the confusion for cloud users [25], and is something which must be addressed in the future.

**Data Segregation**

Multi-tenancy is one of the major characteristics of cloud computing. As a result of multi-tenancy, multiple users can store their data using the applications provided by SaaS. In such a situation, data of various users will reside at the same location. Intrusion of data of one user by another becomes possible in this environment. This intrusion can be done either by hacking through the loop holes in the application or by injecting client code into the SaaS system. A client can write a masked code and inject into the application. If the application executes this code without verification, then there is a high potential of intrusion into other's data. A SaaS model should therefore ensure a clear boundary for each user's data. The boundary must be ensured not only at the physical level but also at the application level. The service should be intelligent enough to segregate the data from different users. A malicious user can use application vulnerabilities to hand- craft parameters that bypass security checks and access sensitive data of other tenants.

**Data Access**

Data access issue is mainly related to security policies provided to the users while accessing the data. In a typical scenario, a small business organization can use a cloud provided by some other provider for carrying out its business processes. This organization will have its own security policies based on which each employee can have access to a particular set of data. The security policies may entitle some considerations, wherein, some of the employees are not given access to certain amount of data [26]. These security policies must be adhered by the cloud to avoid intrusion of data by unauthorized users [27].The SaaS model must be flexible enough to incorporate the specific policies put forward by the organization. The model must also be able to provide organizational boundary within the cloud because multiple organization will be deploying their business processes within a single cloud environment.

**Web application security**

SaaS is software deployed over the internet and/or is deployed to run behind a firewall in local area network or personal computer. The key characteristics include Network-based access to, and management of, commercially available software and managing activities from central locations rather than at each customer's site, enabling customers to access application remotely via the Web. SaaS application development may use various types of software components and frameworks. These tools can reduce time-to-market and the cost of converting a traditional on-premise software product or building and deploying a new SaaS solution. Examples include components for subscription management, grid computing software, web application frameworks and complete SaaS platform products. One of the ''must-have'' requirements for a SaaS application is that it has to be used and managed over the web [28]. The software which is provided as a service resides in the cloud without tying up with the actual users. This allows improvising the software without inconveniencing the user. Security holes in the web applications thus create a vulnerability to the SaaS application [29]. In this scenario, the vulnerability can potentially have detrimental impact on all of the customers using the cloud. The challenge with SaaS security is not any different than with that of any other web application technology. However one of the problems is that traditional network security solutions such as network firewalls, network intrusion detection and prevention systems (IDS & IPS), do not adequately address this problem. Web applications introduce new security risks that cannot effectively be defended against at the network level, and do require application level defenses. The Open Web Application Security Project [30] has provided the ten most critical web applications security threats.

**Data breaches**

Since data from various users and business organizations lie together in a cloud environment, breaching into the cloud environment will potentially attack the data of all the users. Thus, the cloud becomes a high value target [31]. In the Verizon Business breach report blog it has been stated that external criminals pose the greatest threat (73 percent), but achieve the least impact (30,000 compromised records), resulting in a Virtualization vulnerability [32].

**Backup**

The traditional backup methods used with earlier applications and data centers that were primarily designed for web and consumer applications, are not optimally designed for the applications running in the cloud. The SaaS vendor needs to ensure that all sensitive enterprise data is regularly backed up to facilitate quick recovery in case of disasters. Also the use of strong encryption schemes to protect the backup data is recommended to prevent accidental leakage of sensitive information. In the case of cloud vendors such as Amazon, the data at rest in S3 is not encrypted by default. The users need to separately encrypt their data and backups so that it cannot be accessed or tampered with by unauthorized parties.

**Identity management and sign-on process**

Identity management (IdM) or ID management is an area that deals with identifying individuals in a system and controlling the access to the resources in that system by placing restrictions on the established identities. This area is considered as one of the biggest challenges in information security. When a SaaS provider has to know how to control who has access to what systems within the enterprise it becomes all the more challenging task. In such scenarios the provisioning and de-provisioning of the users in the cloud becomes very crucial.

## 3. CURRENT SECURITY SOLUTIONS

There are several research works happening in the area of cloud security. Several groups and organization are interested in developing security solutions and standards for the cloud. The Cloud Security Alliance (CSA) is gathering solution providers, non- profits and individuals to enter into discussion about the current and future best practices for information assurance in the cloud [11]. The Cloud Standards website collects and coordinates information about cloud-related standards under development by the groups. The Open Web Application Security Project (OWASP) maintains list of top vulnerabilities to cloud-based or SaaS models which is updated as the threat landscape changes [30]. The Open Grid Forum publishes documents to containing security and infrastructural specifications and information for grid computing developers and researchers [22]. The best security solution for SaaS applications is to develop a development framework that has tough security architecture. Tsai W [33], put forth a four-tier framework for web-based development that though seems interesting, only implies a security facet in the process. In his work, Berre [34] has suggested a road map towards cloud-centric development, and the X10 language is one of the ways to achieve better use of cloud capabilities of massive parallel processing and concurrency [35].Another approach is resource isolation to ensure security of data during processing, by isolating the processor caches in virtual machines, and isolating those virtual caches from the hypervisor cache [36]. One simple solution, for UK businesses is to simply use in-house ''private clouds'' [37] .Pearson highlighted that the current lack of transparency is preventing many users from reaping the true benefits of the cloud [38].

For Identity and access management in the SaaS,[12] has issued an Identity and Access Management Guidance which provides a list of recommended best practices to assure identities and secure access management. Resource Locality and Data Segregation are the two key security challenges on which not much information is available in the existing literature, which necessitates that this can be further taken up for research.

Table 1: Current solutions available for securing SaaS

| S.No. | Security Areas | Current /Possible Solutions |
|---|---|---|
| 1 | Authentication and authorization | • Open Authorization [5] <br> • Two Factor Authentication [6] <br> • OAuth [39] |
| 2 | Availability | • Data Dispersion[8] |
| 3 | Data confidentiality | • Attribute based Proxy Re-Encryption [40] |
| 4 | Virtual Machine Security | • Reconfigurable distributed virtual machine[41] <br> • Survey on Virtual machine Security [42] |
| 5 | Information Security | • Information Security Risk Management Framework [43] |
| 6 | Network Security | • Network Security for virtual machines [44] <br> • Network Security Sandbox[45] |
| 7 | Cloud standards | • IEEE Cloud Computing Standard Study Group [18] <br> • ITU Cloud Computing Focus Group [19], <br> • Cloud Security Alliance (CSA) |
| 8 | Data Access | • Multi-user access policies [46]. <br> • Data Access Management [47]. |
| 9 | Web application security | • Web Application Scanners[48] |
| 10 | Data breaches | |
| 11 | Backup | • Agentless Method for data Backup and Recovery [49] |
| 12 | Identity management and sign-on process | • CSA's Identity and Access Management Guidance [12] |

## 4.CONCLUSION

Though there are numerous advantages in using a cloud-based system, there are yet many practical issues which have to be sorted. Cloud computing is a disruptive technology with profound implications not only for Internet services but also for the IT sector as a whole. Still, several outstanding issues exist, particularly related to service-level agreements (SLA), security and privacy, and power efficiency. As described in the paper, currently security has lot of loose ends which scares away several potential users. Until a proper security module is not in place, potential users will not be able to leverage the true benefits of this technology. This security module should cater to all the issues arising from all directions of the cloud. Every element in the cloud should be analyzed at both the macro and micro level and subsequently an integrated solution must be designed and deployed in the cloud to attract and retain the potential consumers. Until then, cloud environment will remain cloudy. In a cloud, where there are heterogeneous systems having a variation in their asset value, a single security system would be too costly for certain applications and if there is less security then the vulnerability factor of some applications like financial and military applications will shoot up. On the other side, if the cloud has a common security methodology in place, it will be a high value asset target for hackers because of the fact that hacking the security system will make the entire cloud vulnerable to attack.

In this paper an overview of cloud computing service delivery model, SaaS along with the security challenges , including both the traditional and cloud specific security challenges ,associated with the model has been presented A number of new challenges that is inherently connected to the new cloud paradigm has also been deliberated in the paper. As secure data storage in cloud environment is a significant concern which prevents many users from using the cloud, a practical solution to provide security and privacy for user data, when it is located in a public cloud, was also discussed in this paper. The need for further work on various security mechanisms has also been highlighted, in order to provide transparent services that can be trusted by all users.

## Authors

**Rashmi**
Rashmi is a researcher at Birla Institute of Technology. Her current research focus is on Cloud Computing and securing Cloud environments. She holds her B.E. degree in Information Technology from Pune University, India in 2005 and M.Tech. In Computer Science in 2008.She was awarded "Gold Medal" for her overall academic excellence. She has also worked as a Software Engineer in reputed IT firms before joining academic career. She can be contacted at: *rashmirrai@gmail.com*

**Dr.G.Sahoo**
G. Sahoo received his MSc in Mathematics from Utkal University in the year 1980 and PhD in the Area of Computational Mathematics from Indian Institute of Technology, Kharagpur in the year 1987. He has been associated with Birla Institute of Technology, Mesra, Ranchi, India since 1988, and currently, he is working as a Professor and Head in the Department of Information Technology. His research interest includes Theoretical Computer Science, Parallel and Distributed Computing, Cloud Computing, Evolutionary Computing, Information Security, Image Processing and Pattern Recognition.

**Dr.Shabana Mehfuz**
Shabana Mehfuz received her B.Tech degree in Electrical Engineering from Jamia Millia Islamia, India in 1996 and M.Tech degree in Computer technology from IIT Delhi in 2003 and PhD in Computer Engineering from Jamia Millia Islamia in 2008. She is currently working as Associate Professor at the Department of Electrical Engineering. Her research interest includes Computational Intelligence and Computer Networks.